\documentclass{aa}
\usepackage{graphicx}
\usepackage{txfonts}
\usepackage{natbib}
\bibpunct[; ]{(}{)}{,}{a}{}{,}
\def\asec{\ifmmode ^{\prime\prime}\else$^{\prime\prime}$\fi}

\def\grad{$^\circ$}

\def\degs{\ifmmode ^{\circ}\else$^{\circ}$\fi}
\def\amin{\ifmmode ^{\prime}\else$^{\prime}$\fi}
\def\asec{\ifmmode ^{\prime\prime}\else$^{\prime\prime}$\fi}
\def\farcs{\hbox{$.\!\!^{\prime\prime}$}}  

\def\degs{\ifmmode ^{\circ}\else$^{\circ}$\fi}
\def\amin{\ifmmode ^{\prime}\else$^{\prime}$\fi}
\def\farcm{\hbox{$.\mkern-4mu^\prime$}}
\def\eqalign#1{\null\,\vcenter{\openup1\jot \m@th
   \ialign{\strut\hfil$\displaystyle{##}$&$\displaystyle{{}##}$\hfil
   \crcr#1\crcr}}\,}
\begin{document}
   \title{ 
   The G292.0+1.8 pulsar wind nebula in the mid-infrared
}
\author{D.A. Zyuzin\inst{1,2}
 \and  A.A.~Danilenko\inst{1}  \and S.V.~Zharikov\inst{3} \and Yu.A.~Shibanov\inst{1} }
\offprints{D.A. Zyuzin, \\  \email{ zyuzia@mail.ru}}
\institute{ Ioffe Physical Technical Institute, Politekhnicheskaya
26,
St. Petersburg, 194021, Russia \and Academical Physical Techonological
University,  Khlopina 2-8,   St. Petersburg, 194021, Russia 
\and Observatorio Astron\'{o}mico Nacional SPM, Instituto de Astronom\'{i}a, UNAM, Ensenada, BC, Mexico}
  \abstract
   {G292.0+1.8 is a Cas A-like supernova remnant that contains the young
pulsar  PSR J1124-5916 powering a compact torus-like pulsar wind nebula 
visible in X-rays. A likely counterpart to the nebula  
has been detected   in the optical  $VRI$ bands.  
}
   { To confirm the  counterpart candidate  nature, we examined  archival mid-infrared data obtained with the Spitzer Space Telescope.}
   { Broad-band images  taken  at 4.5, 8, 24, and 70 $\mu$m were analyzed and compared  with   available
  optical and  X-ray data.}
   {  The extended  counterpart candidate is firmly detected  in  the 4.5 and 8 $\mu$m  bands. 
      It is brighter and more extended in the bands than in the optical, and 
      its position and morphology  
      agree well with the coordinates and morphology of the torus-like pulsar wind nebula  
      in X-rays.  The source is not visible in  24 and 70 $\mu$m  images, which are dominated by   bright  
      emission from the remnant shell and filaments.  
    We compiled the infrared fluxes  of the nebula,  which probably  
    contains a contribution from an unresolved pulsar in its center, 
    with the  optical  and X-ray  data. The resulting unabsorbed multiwavelength spectrum is described by power 
    laws  of significantly steeper slope  in the infrared-optical than in  X-rays, implying   a double-knee spectral break between 
    the optical and X-rays. The 24 and 70 $\mu$m flux upper limits suggest a second break  and  a flatter spectrum  
    at the long wavelength limit.  These features are common to 
two other pulsar wind nebulae  associated with the  remnants B0540-69.3 and 3C 58 and observed in all three ranges. }
    {The position, morphology, and spectral properties of the detected source
allow us to comfirm   that it is  the infrared-optical counterpart to both the pulsar and its wind nebula system 
in the G292.0+1.8 supernova remnant.} 

\authorrunning{D.A. Zyuzin et al. }
\titlerunning{
   The G292.0+1.8 pulsar wind nebula in the mid-IR. }
   \maketitle
%
%
%
\section{Introduction}
\label{sec1}
Infrared (IR) observations can   provide  valuable 
information  about  the poorly understood physical processes responsible for
multiwavelength radiation from  rotation-powered neutron stars  
and  pulsar wind nebulae (PWNe). However, until now only a few of
these objects have been detected   in the IR. The  most well studied
Crab-pulsar and its torus-like PWN exhibit   almost a flat
spectrum in the IR \citep{tem06, tem09},  while some bright
structures of the PWN, such as a knot south of the pulsar, exhibit 
 a steep flux increase towards longer wavelengths
\citep{sol03, sol09}.   In contrast to  the emission of the Crab-pulsar itself,  the 
emission from three older pulsars   detected in the near-IR (Vela, B0656+14, Geminga)  show   
a flux increase with wavelength  \citep{shib03, shib06}. 
 
 Two  young  Crab-like pulsar+PWN systems associated with supernova remnants (SNRs) 
 B0540-69.3   and   3C~58,    observed in the mid-IR with the Spitzer  telescope,   exhibited an even  steeper
 flux increase  towards the IR \citep{wil08, sla08, shib08}. 
 Their optical and IR data   can be described by a single power law, whose slope and
 intensity  differ significantly from those observed in X-rays suggesting a double-knee 
 break in the multiwavelength spectra of both systems between the optical
 and X-rays \citep{ser04, shib08}. The  emission from the
  Crab   contains only one break within the same range.
  More complex energy distributions of radiating relativistic particles
   in the two former systems may be resposible   
   for this difference. IR observations  of other pulsar-PWN systems can help us to understand  whether  
   the double-knee break is a more common
 feature  of these systems than  the single break seen in the Crab.
 
 The young  \object{PSR J1124-5916}  was only recently  discovered in
the radio  \citep{Camilo}  and X-rays \citep{Hughes1,Hughes2}.
It is associated  with   \object{SNR  G292.0+1.8} (MSH 11-54), which is  the  third  oxygen-rich  
SNR to be detected in the Galaxy after Cas A and  Puppis A.  
In a similar way to  the  Crab-pulsar,  PSR J1124-5916  powers a torus-like  X-ray PWN with a jet \citep{Hughes1,safi02,Hughes2,park07}.
The  characteristic age of the pulsar,  $\tau$ $\approx$ 2900 yr,  which is
consistent with   2700--3700 yr age of the SNR \citep{Camilo,chev05}, and  its spin-down luminosity,  
$\dot{E}$ $\approx$1.2$\times10^{37}$ ergs~s$^{-1}$, rank this pulsar as the sixth youngest and the eighth most energetic
among the all rotation-powered pulsars  known.    
Deep  optical images    obtained last year  
with the VLT detected  a likely candidate to the optical counterpart of the pulsar+PWN system in $VRI$  bands \citep{zhar08-2}.
The major axis of an elliptically shaped counterpart candidate  coincides with the major
axis of the torus-like PWN, which is seen in X-rays almost edge-on.
Smaller source extents  suggest that only the brightest inner part
of the X-ray PWN containing the pulsar was detected in the optical range.

Using  Spitzer archival images of  the G292.0+1.8 field,   
 we report the detection of the same source in the  mid-IR.
Similar source morphology to that in X-rays and a single power law  spectral energy distribution 
in the optical and IR show that it is indeed the optical-mid-IR counterpart of the  G292.0+1.8 pulsar+PWN system.
The IR observations are described in Sect.~\ref{sec2},  the results 
are presented and discussed in  Sects. \ref{sec3} and \ref{sec4}. 
\section{Observations}
\label{sec2}
G292.0+1.8 was imaged with the Spitzer telescope at several observing sets 
from  March 13 to  April 15 2008  using the Infrared Array Camera (IRAC)  and Multiband Imaging
Photometer for Spitzer (MIPS)\footnote {see http://ssc.spitzer.caltech.edu}.  The data\footnote{
PROGID 40583,  PI P. Ghavamian} were retrieved
from the Spitzer archive. Various parts of the SNR were detected 
in all four IRAC channels, although its central part, containing the pulsar,  was imaged
in only  the second and  fourth bands with  effective wavelengths of
 4.5 $\mu$m   and 8.0 $\mu$m, respectively.
 We used the  IRAC  post-BCD calibrated mosaic images with an effective image
 scale of 1\farcs2 per pixel and field of view (FOV) of 7\farcm5$\times$7\farcm5.
 Two separate sets of observations were obtained with
 the MIPS at the effective wavelengths of  24 $\mu$m   and  70 $\mu$m to image the entire SNR.
 We regenerated the MIPS  post-BCD
 calibrated mosaic images  from  respective  BCD
 data using the MOPEX\footnote{http://ssc.spitzer.caltech.edu/postbcd/mopex.html} tool. The resulting image scales are 2\farcs4 
 per pixel for the first, and  4\farcs0 for the second band with the FOV of 13\farcm3$\times$12\farcm8 for both bands.
  The effective exposures were $\approx$3500 s per pixel for both IRAC bands,  $\approx$500 s for 24 $\mu$m,  and
 $\approx$380 s  for 70 $\mu$m MIPS bands.
 \section{Results}
 \label{sec3}
 \subsection{Identification  of the  pulsar/PWN  counterpart candidate }
 As a first step, to check the Spitzer pointing accuracy, we applied  astrometric referencing
to the IRAC images. In the  IRAC FOV, eight  unsaturated isolated
astrometric standards from the USNO-B1 catalog were selected   as
reference points. Their catalog  and image positional uncertainties
for both coordinates  were $\la$0\farcs2 and $\la$0\farcs5,
respectively.   IRAF  {\sl ccmap/cctran} tasks  were used to find
plate solutions. Formal {\sl rms} uncertainties of  the astrometric
fit were $\la$0\farcs2  with maximal  residuals of $\la$0\farcs5 for
both coordinates and both IRAC bands. After astrometric
transformations the shifts between the original and transformed
images were $\la$0\farcs4 (or less than 0.3 of the pixel scale),
ensuring the almost  perfect pointing accuracy of the Spitzer observations.
Combining  all uncertainties,  a conservative estimate of 1$\sigma$
referencing uncertainty is  $\la$0\farcs5 (or less than 0.4 of the IRAC pixel scale)  in both coordinates for both  bands.
This is  about twice as poor as the astrometric uncertainties estimated by \citet{zhar08-2} for  the deepest available Chandra/ACIS-I
X-ray ($\la$0\farcs27) and VLT/FORS2 ($\la$0\farcs15)  optical images of G292.0+1.8. Nevertheless, this  is sufficient  
 to identify positionally the objects in the mid-IR, optical, and X-rays   on a subarcsecond  accuracy level.

\begin{figure*}[t]
\setlength{\unitlength}{1mm}
\resizebox{12.cm}{!}{
\begin{picture}(120,105)(0,0)
\put (10,50) {\includegraphics[width=16 cm, clip]{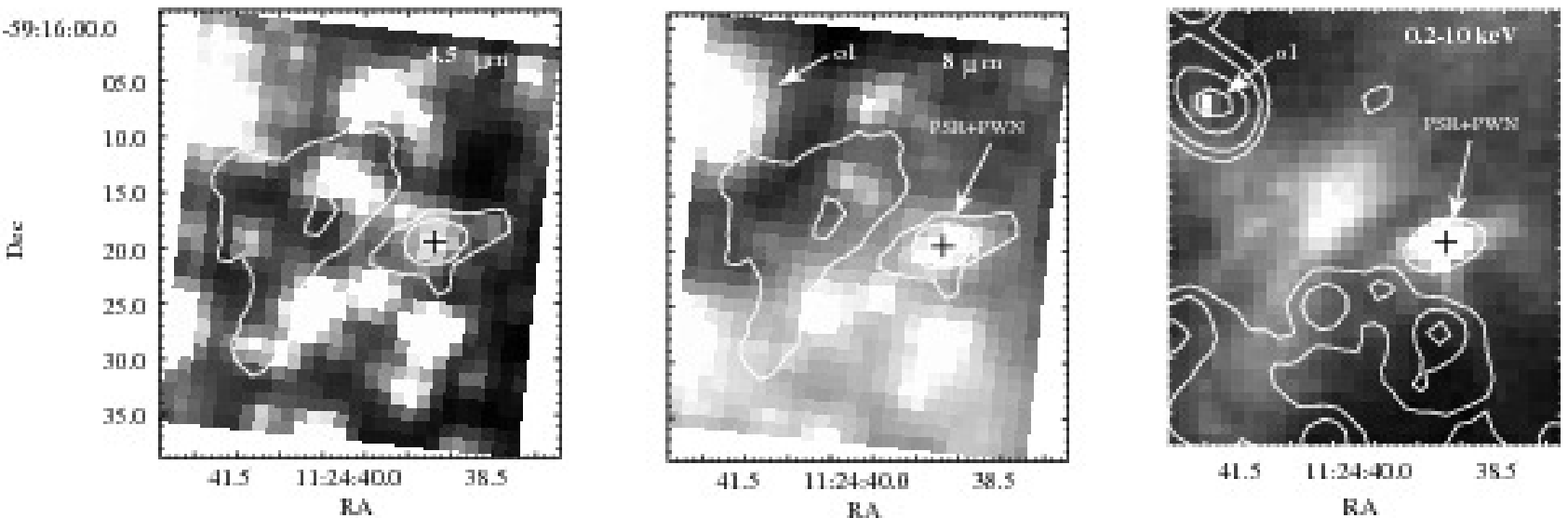}}
\put (10,0) {\includegraphics[width=16 cm,  clip]{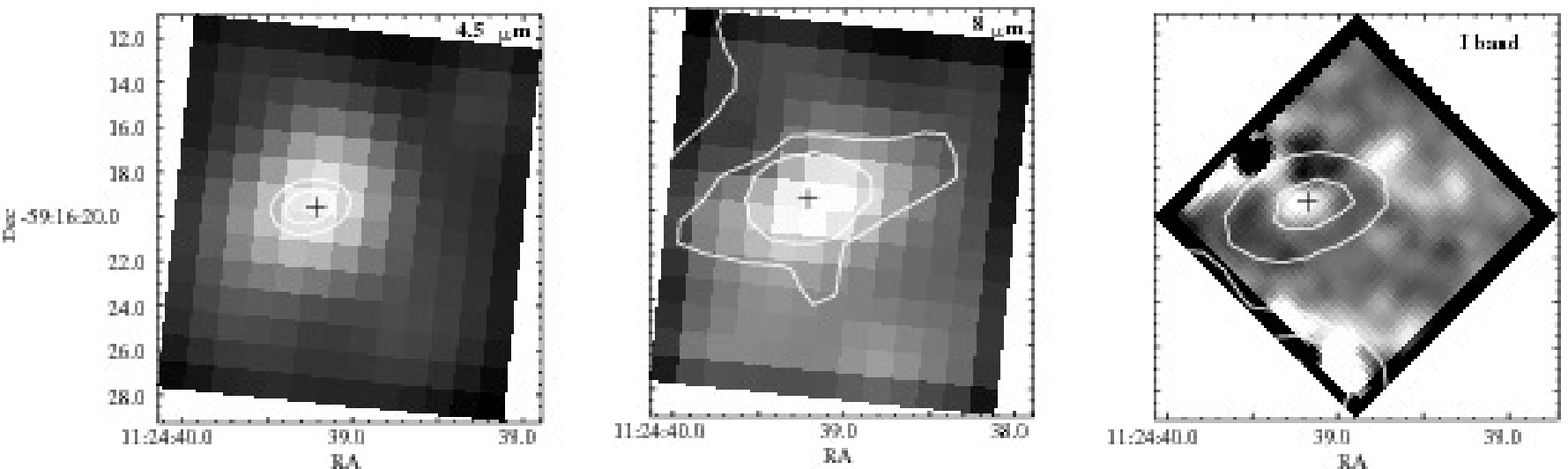}}

\end{picture}}
  \caption{{\sl Top row:} central  $\sim$35\asec$\times$40\asec~fragment of G292.0+1.8, containing PSR J1124-5916 and its PWN 
  as seen with the Spitzer/IRAC and Chandra/ACIS-I at 4.5 $\mu$m and 8 $\mu$m, and in 0.2--10 keV X-ray range, respectively.  
   X-ray position of the pulsar is  marked by the cross.
  Contours from the X-ray image  are  overlaid on the mid-IR images, while in X-rays the contours are
  from  the 8 $\mu$m image.  Arrows  indicate 
   the pulsar+PWN system and a background object o1  in X-rays and their mid-IR counterparts.
 {\sl Bottom row:}  zoomed fragments of the same mid-IR images showing  a $\sim$17\asec$\times$17\asec~ region around the pulsar  
 and respective  $I$ band optical image obtained with the VLT/FORS2. Background stars were  subtracted.
 Contours of the optical pulsar+PWN counterpart candidate are overlaid on  4.5 $\mu$m image. Contours on the $I$ image 
 are from 8 $\mu$m, while  on the 8 $\mu$m image the X-ray contours are overlaid.
 The images are smoothed  with a Gaussian kernel of two-three pixels.  The coordinates and morphology of the source seen 
 at the pulsar position  in the optical and mid-IR suggest that it is  the counterpart of  the X-ray pulsar+PWN system.
 }
 \label{fig1}
\end{figure*}

The region  containing the pulsar is shown in Fig.~\ref{fig1}, where we compare
the IRAC mid-IR images with the X-ray and optical images obtained with Chandra/ACIS-I
and VLT/FORS2  \citep{zhar08-2}.
In both IRAC  bands  at the  pulsar  position,   marked by a cross,
we detect  a faint extended source.  The FWHM of the IR point spread function 
varies from 2\farcs1 to 2\farcs5, and the extended structure
of the source,  which subtends up to $\approx$10\farcs5,  is clearly resolved.
In the 8 $\mu$m  image, which is less contaminated by background stars, the object has an elliptical shape, as can  
also be seen from its contours overlaid on the X-ray and optical images\footnote{ In both images, the contours are identical,     
but the internal  IR contour  seen in the optical falls inside the  X-ray PWN and  is not visible  
on chosen  scale levels  in the X-ray image.}.  
The coordinates of the source center, derived from  the source surface brightness fit  in 8 $\mu$m band  by elliptical isophotes, are
RA$\approx$11:24:39.129 and  Dec$\approx$-59:16:19.51. Accounting for the astrometric uncertainties, this is  in  good agreement
with  the X-ray position of the pulsar,  RA$\approx$11:24:39.183 and Dec$\approx$-59:16:19.41, and with the central position  
of the optical pulsar/PWN counterpart candidate, RA$\approx$11:24:39.216 and Dec$\approx$-59:16:19.60 \citep{zhar08-2}.
The source ellipticity and position angle are   compatible with the structure of the brightest inner part
of the X-ray PWN, which has been interpreted as a Crab-like PWN torus    seen almost edge-on \citep{Hughes2}.

The position and morphology of the detected mid-IR source allow us 
to consider  whether it  is a true counterpart to the pulsar+PWN system.  
A faint jet-like structure extended south of the  torus
and a bright SNR filament visible N-E of the PWN in X-rays are not
detected in the mid-IR. In the  frames  presented in Fig.~\ref{fig1}, 
we note   only one additional cross-identification, an object o1,
showing a point-like  spatial profile in X-rays but  a non-stellar
structure in the IR. This may be a background galaxy. 
Star-subtracted optical and mid-IR images (bottom row panels of
Fig.~\ref{fig1}) show that at the pulsar position we see the same
extended elliptical object in both ranges. This confirms the reality of its
optical detection  in the region that is extremely  crowded by  background  objects \citep{zhar08-2}. 
 Most  nearby stars contaminating the
source flux in the optical becomes fainter  in the IR, while the
brightness of the source itself does not decrease, enabling its
detection in the mid-IR at a higher significance level. As in the
optical, its surface brightness distribution  reaches a peak at the pulsar
position. We cannot resolve any point-like object there, but the peak 
is probably associated with the contribution  from the pulsar.  Some blurring is caused by
a lower   mid-IR spatial resolution  compared to the optical one.  However,  the object sizes  along its major axis 
corresponding to  $\ga$90\% of the enclosed flux are significantly larger in the IR,  $\approx$10\farcs5, 
than in  the optical, $\approx$3\asec.   This suggests   that  the source  becomes brighter with 
wavelength  (see below Sect.~3.2 and 3.3), revealing its fainter outer regions.  At  8 $\mu$m,   
it  becomes comparable in   size and shape  to the  torus part of the PWN seen in X-rays.

Our counterpart candidate is not detected   in the MIPS bands, where
the emission from filaments and  the outer shell  of the SNR  strongly dominates  over other sources
in contrast to  the shorter wavelength IRAC bands.
Some faint structure may be  present at the pulsar position,
particularly in the 24  $\mu$m image.   However,  the poorer spatial
resolution of  MIPS  does not allow us to determine, whether it is related
to the  candidate or to a  faint peninsula-like part of a bright  nearby  filament of the remnant.
\subsection{Photometry}
Elliptical aperture photometry  of the detected source
was performed 
on the star-subtracted IRAC images
using IRAF {\sl polyphot}  tasks in accordance with
prescriptions and zeropoints given in the IRAC Observers Manual\footnote{see, e.g.,
http://ssc.spitzer.caltech.edu/archanaly/quickphot.html}. The elliptical apertures were obtained
from the source surface brightness fit  at  8 $\mu$m   by elliptical isophotes using IRAF {\sl isophot} tasks.
The   ellipticities varied with the ellipse sizes  in a range  of 0.4--0.5 and the semimajor axis
of the outer isophote was found to be $\la$6 pixels (7\farcs2).
The annulus for backgrounds was 8--23 pixels centered on the pulsar position.
 A typical semi-major radius, where the curves of growth saturate,
 was about 5 pixels for both IRAC bands. We reiterated  the star subtraction process,
varied the background region, and used circular apertures. The differences in the magnitudes
obtained were similar to  the measurement statistical errors and they were
included in the resulting uncertainties.
A few percent of the extended source aperture corrections were insignificant  compared to  
 the resulting $\approx$20\%--30\%  flux error budget.

  The 3$\sigma$    upper limits to the source magnitudes in the MIPS bands were estimated
 using  the  standard deviations  and inherent flux variations in the possible faint part of the remnant  filament   
 within the same elliptical apertures as in the IRAC bands. Both ways provided  similar values.   
 For  70 $\mu$m,  we used a post-BCD filtered calibrated mosaic image generated  by us
 with the MOPEX tool. 
  The magnitudes  were converted into fluxes in physical units and
 the results are summarized in Table~\ref{t:phot}.  Our 70 $\mu$m limit  is slightly above 
 the 5$\sigma$ flux confusion limit of 1500 $\mu$J estimated  for this MIPS band by \citet{Frayer06}.   
\begin{table}[t]
\caption{Observed magnitudes and fluxes for the presumed
infrared/optical  pulsar/PWN counterpart of J1124-5916, and
de-reddened fluxes for the $A_V$ range of 1.86-2.10.}
\begin{tabular}{llll}
\hline\hline
  $\lambda_{eff}$(band)           &     Mag.                        &  log Flux                                 & log Flux                       \\
                                                      &  observed$^a$            &  observed$^a$                       &  de-reddened$^a$                    \\
($\mu$m)                       & (mag)                           & ($\mu$J)                                  & ($\mu$J)    \\
\hline
                                        &           infrared              &         Spitzer                            &                              \\
 \hline
  70                                 & $\ga$6.8                         &$\la$3.2                                   &$\la$3.2   \\
 24                                 &  $\ga$10.2                      &$\la$2.8                                   &$\la$2.85    \\
 8.0                               & 14.2(3)                           &2.13(16)                                    & 2.18($^{+16}_{-17}$)    \\
 4.5                              & 15.9(4)                            &1.90(21)                                    &  1.95($^{+21}_{-22}$)  \\
 \hline
                                     &  optical                                 &                     VLT                  &                                 \\
\hline
 0.77($I$)                   & 23.12(13)                      & 0.13(5)                                        &  0.66($^{+6}_{-10}$)    \\
 0.66($R$)                  & 24.12(13)                    & -0.17(5)                                       &  0.51($^{+7}_{{\bf-12}}$)         \\
 0.55($V$)                 & 24.29(13)                    & -0.16(5)                                        & 0.66($^{+7}_{-13}$)          \\
\hline
\end{tabular}
\label{t:phot}
\begin{tabular}{ll}
$^a$~numbers in brackets are 1$\sigma$ uncertainties referring  & \\
to last significant digits quoted & \\
\end{tabular}
\end{table}
\subsection{Multiwavelength spectrum}
Using the mid-IR, optical, and X-ray data, we  compiled a tentative
multiwavelength spectrum of   the pulsar/PWN system. We
dereddened the observed optical/IR fluxes using a most plausible
interstellar extinction range  1.86$\la$$A_V$$\la$2.10   discussed
in detail  by  \citet{zhar08-2}. Standard  extinction
curve (Cardelli et al. 1989) and  average  $A_{\lambda}/A_K$
ratios provided especially for the IRAC bands \citep{indeb05,Flaherty07}
 were applied.  The results are presented in Table  \ref{t:phot}, where we also include the optical data obtained
 by \citet{zhar08-2}.
 
To compare the fluxes in different ranges from the same physical region, 
we extracted  the pulsar/PWN X-ray spectrum from the archival Chandra/ACIS-I data\footnote{Obs 6677-6680, 8221, 8447,
530 ks exposure, PI S. Park}  using  the same elliptical aperture, as for the mid-IR photometry of the detected source, 
with the semimajor axis of 5\farcs2,
ellipticity  $\approx$0.5, and the PA $\approx$-70\grad.
An absorbed power law spectral model provided a statistically acceptable fit
 with a photon spectral index $\Gamma$=1.85$\pm$0.02,
absorbing column density $N_{H}$=(3.4$\pm$0.1$)\times$10$^{21}$~cm$^{-2}$, and
normalization constant $C$=(2.4$\pm$0.1)$\times$10$^{-4}$~photons~cm$^{-2}$~s$^{-1}$~keV$^{-1}$.
The fit had $\chi^2$=1.1 per degree of freedom ({\sl dof}) and the unabsorbed integral flux
was 1.0$\times10^{-12}$~erg~cm$^{-2}$~s$^{-1}$ in 0.3--10 keV  range.
The resulting $\Gamma$ and $N_{H}$  are compatible with those  of 
\citet{zhar08-2}, while the normalization constant and integral flux are about twice as high. This is
 because of 
 a larger extraction aperture than  in their work, where the aperture was chosen to fit  the smaller
 source extent in the optical range.  The  optical fluxes  are  unchanged  by the aperture increase
 since the likely counterpart  is more compact in the optical than in the IR.

\begin{figure}[t]
 \setlength{\unitlength}{1mm}
\includegraphics[width=63mm, angle=-90, clip=]{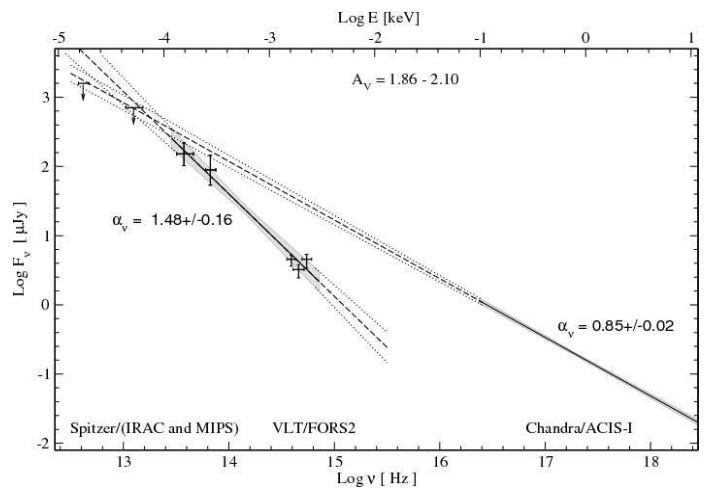}
 \caption {Tentative unabsorbed multiwavelength spectrum for the inner part of the torus region of the G292.0+1.8
 pulsar/PWN system compiled from the data obtained with different instruments,   
 as indicated in the plot. Errorbars of the dereddened optical and mid-IR fluxes  include the $A_V$ uncertainty range shown at the top.
  The VLT and IRAC   data are  fitted by a single powerlaw, whose spectral index
  $\alpha_{\nu}$ (defined by  $F_{\nu}$ $\propto$ $\nu^{-\alpha_{\nu}}$) is significantly greater than that for X-rays, 
  suggesting a double knee spectral break between the optical and X-rays.  The MIPS flux upper limits are below
  the low frequency extension of this fit and  imply a  second  break at $\lambda$ $\approx$20 $\mu$m.    
  Solid lines  show the best spectral fits, while grey polygons, and dashed and dotted lines are the fit 
  uncertainties and their extensions  outside the  frequency ranges involved in the fits. }
 \label{fig:mw}
 \end{figure}

The resulting unabsorbed spectrum is presented in Fig.~\ref{fig:mw}. The counterpart
candidate spectral energy distribution (SED) in the IR-optical range shows a steep flux increase towards the mid-IR.
 The  fluxes in this range can be fitted by a single power-law with a spectral index  $\alpha_{\nu}$ $\approx$1.5  ($\chi^2$ $\approx$0.5 per {\sl dof}),
 implying a nonthermal nature of the emission.  This  is expected for PWNe,  where
  synchrotron radiation  from  relativistic particles accelerated at the pulsar wind termination shock
   is considered as the main radiative process  responsible for the continuum emission of the nebulae.
   Accounting for possible contamination of the $V$  flux by O III emission from the SNR  \citep{zhar08-2}, 
    we excluded this band from the fit.  This leads to only  a marginal improvement of the fit 
    with insignificant steepening of the spectral slope within the uncertainties shown in the plot.    
   The observed SED  cannot be produced by overlapping  field stars   of  the same brightness.
   Our test measurements of nearby faint  stars show that in  the latter case  the spectral slope would
   be a positive, flat, or curved.

  As seen from Fig.~\ref{fig:mw}, the 
   optical and mid-IR fluxes of the suggested counterpart
  are below the  low frequency  extrapolation of the pulsar/PWN spectrum  in X-rays. This implies
  a double-knee spectral break between the optical and X-rays.  At the same time, the
  flux upper limits in the MIPS bands are below the low frequency extrapolation of the IRAC-optical SED,
  suggesting a second  break near 20 $\mu$m after which the spectrum become flatter
  or goes down with the frequency decrease.  These changes in the spectral index and breaks 
  are similar to those observed for PWNe around young pulsars in  SNRs 3C 58 and B0540-69.3 \citep{ser04, shib08,
  sla08, wil08}. Together with the source morphology, this  is
  a strong evidence that  the detected source is the true  mid-IR and optical counterpart
  of the pulsar/PWN system in  the G292.0+1.8 supernova remnant.
\section{Discussion}
\label{sec4} These broad-band Spitzer observations have allowed us to 
confirm  the  pulsar/PWN counterpart nature of a faint optical
nebulosity detected early at    the PSR J1124-5916 position with the
VLT \citep{zhar08-2}. Its similar source morphology in the mid-IR,    optical, and X-rays, 
as well as its spectral properties make any
alternative interpretation discussed in \cite{zhar08-2} (SNR
filament, faint background spiral galaxy) very unlikely. The mid-IR
data have enabled us to establish the long wavelength SED of the nebula,
which was   very uncertain based only on the optical data, and
conclude that   the multiwavelength spectrum of the  J1124-5916
pulsar/PWN system from the mid-IR to X-rays is similar to the
spectra of  pulsar/PWN systems in the SNRs B0540-69.3 and 3C 58.
Forthcoming high spatial resolution near-IR observations of the
pulsar field will allow us to decrease the rather large uncertainties of
the IR-optical spectral slope  and possible distinguish the pulsar from
the PWN.

Our results increase the number of PWNe identified in the mid-IR  from three to
four,  and show that these objects can be significantly brighter in the IR
than in the optical, making the IR range promising for the study of
the PWNe. The presence of the  double-knee breaks in the spectra  between the optical and X-rays  in
the three  of four torus-like PWNe makes this feature regular and notable.
This is distinct from the Crab-PWN and has to be taken into account in
the modeling   of the PWNe when understanding their structure and emission nature.

\begin{acknowledgements}
     We are grateful to anonymous referee for useful comments improving the paper. 
     The work was partially supported by  
     RFBR (grants 08-02-00837a, 09-02-12080),   NSh-2600.2008.2, CONACYT 48493 and PAPIIT IN101506 projects, 
  and by  the German \emph{DFG\/} project  number Ts~17/2--1. DAZ was supported  by St Petersburg Goverment grant 
  for young scientists (2.3/04-05/008).
\end{acknowledgements}


\begin{thebibliography}{}
\bibitem[Camilo et al.(2002)]{Camilo} Camilo, F., Manchester, R. N., Gaensler, B. M., et al. 2002,
\apj, 567, L71
\bibitem[Cardelli et al.(1989)]{card89}  Cardelli, J. A., Clayton, G. C., Mathis, J. S. 1989, \apj, 345, 245
\bibitem[Chevalier(2005)]{chev05}   Chevalier, R. 2005,  \apj, 619, 839
\bibitem[Fraherty et al.(2007)]{Flaherty07} Flaherty, K. M., Pipher, J.L., Megeath, S.T. et al. 2007, \apj, 663, 1069
\bibitem[Frayer et al.(2006)]{Frayer06}  Frayer D.T. et al. 2006, \apj, 647, L9 
\bibitem[Hughes et al.(2001)]{Hughes1}  Hughes, J. P., Slane, P. O., Burrows, D. N.,
Garmire, G. 2001, \apj, L153
\bibitem[Hughes et al.(2003)]{Hughes2} Hughes, J. P., Slane, P. O., Park, S., et al. 2003,
\apj, 591, L139
\bibitem[Indebetouw et al.(2005)]{indeb05} Indebetouw, R., Mathis, J.J., Babler, B.L. et al. 2005, \apj, 619, 931
\bibitem[Park et al.(2007)]{park07} Park, S., Hughes, J.P., Slane, P.O.  et al. 
2007,  \apj, 670, L121
\bibitem[Safi-Harb \& Gonzalez(2002)]{safi02} Safi-Harb, S. \& Gonzalez, M.E. 2002, in:  "X-rays at Sharp Focus": Chandra Science Symp.
 ASP Comf. Series Vol 262, eds. E. Schlegel and S.D. Vrtilek
\bibitem[Sandberg \& Sollerman(2009)]{sol09}   Sandberg, A., Sollerman, J.   2009, A\&A, 504, 525 
\bibitem[Serafimovich et al.(2004)]{ser04}    Serafimovich, N., Shibanov, Yu.A., Lundqvist, P., Sollerman, J.
 2004,   A\&A, 425, 1041
\bibitem[Shibanov et al.(2003)]{shib03} Shibanov, Yu. A., Koptsevich, A. B., Sollerman, J., Lundqvist, P. 2003, A\&A,  406, 645
\bibitem[Shibanov et al.(2006)]{shib06} Shibanov, Y. A., Zharikov, S. V., Komarova, V. N. et al. 2006  , A\&A, 448, 313
\bibitem[Shibanov et al.(2008)]{shib08} Shibanov, Yu.A., Lundqvist, N., Lundqvist, P. et al. 
2008, A\&A,  486, 273
\bibitem[Slane et al.(2008)]{sla08} Slane, P., Helfand, D. J., Reynolds. S.P., et al. 2008, \apj, 676, L33
\bibitem[Sollerman(2003)]{sol03}   Sollerman, J.  2003,  A\&A, 46, 639
\bibitem[Temim et al.(2006)]{tem06}   Temim, T., Gehrz, R. D., Woodward, C. E.  et al. 2006, \aj, 132, 1610
\bibitem[Temim et al.(2009)]{tem09}   Temim, T., Gehrz, R. D., Woodward, C. E.  et al. 2009, \aj, 137, 5155
\bibitem[Williams et al.(2008)]{wil08} Williams, B. J., Borkowski, K. J., Reynolds, S. P. et al. 2008, \apj, 687, 1054
\bibitem[Zharikov et al.(2008)]{zhar08-2} Zharikov, S., Shibanov, Yu., Zyuzin D.A. et al., 
2008, A\&A, 492, 805
\end{thebibliography}
\end{document}